\documentclass{jpsj2}
\title{Neutron-Scattering Study of the Localized Mode in $\beta$-Pyrochlore Superconductors AOs$_{2}$O$_{6}$}

\author{Kenzo \textsc{Sasai}\thanks{E-mail address: sasai@issp.u-tokyo.ac.jp}, Kazuma \textsc{Hirota}, Yohei \textsc{Nagao}, Shigeki \textsc{Yonezawa} and Zenji \textsc{Hiroi}}

\inst{Institute for Solid State Physics, University of Tokyo, Kashiwa, Chiba 277-8581}

\abst{Inelastic neutron-scattering and neutron powder-diffraction experiments were carried out to investigate the localized mode, proposed from various bulk measurements, in the $\beta$-pyrochlore AOs$_{2}$O$_{6}$ (A=K, Rb, Cs). The localized mode was identified in all three compounds as well as in another $\beta$-pyrochlore CsW$_{2}$O$_{6}$. The anharmonicity of the mode is weak in RbOs$_{2}$O$_{6}$ and CsOs$_{2}$O$_{6}$ but substantial in KOs$_{2}$O$_{6}$.}

\kword{pyrochlore, superconductivity, localized mode, rattling motion, neutron scattering, anharmonicity}

\begin{document}
\maketitle

\section{Introduction}
Materials with oversized cages have attracted much attention recently because the vibration of a metal ion in such a cage appears to give rise to unusual phenomena.~\cite{sales96, keppens98,paschen01}$^{)}$ This vibration is different from that of ions in ordinary systems, in which an ion vibrates coherently with surrounding ions. However, an ion in an oversized cage may vibrate incoherently or independently because of the weak bonding. This incoherent vibration is called the `localized mode' or `rattling motion,' and is observed in filled skutterudite~\cite{sales96, keppens98}$^{)}$ and clathrate compounds.~\cite{paschen01}$^{)}$ The localized mode is expected to scatter heat-transporting acoustic phonons, resulting in a decrease in the lattice thermal conductivity. This property has great potential for use in thermoelectric devices.

\begin{figure}[tb]
\begin{center}
\includegraphics[scale=0.4]{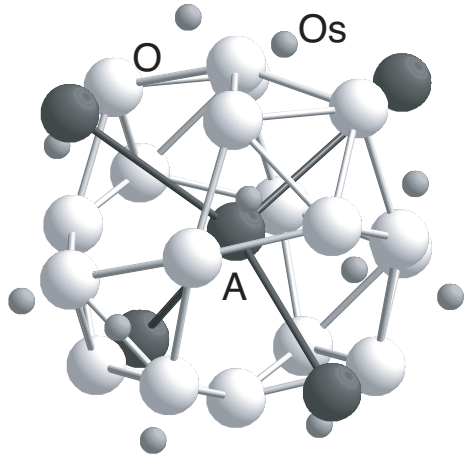}
\end{center}
\caption{Structure of $\beta$-pyrochlore superconductors. Dark spheres indicate alkali-metal ions; grey, Os; white, O.}
\label{structure.f}
\end{figure}

Recently discovered $\beta$-pyrochlore superconductors AOs$_{2}$O$_{6}$ (A=K, Rb, and Cs) contain oversized cages (see Fig. \ref{structure.f}).~\cite{yone04k, yone04rb, yone04cs, hiroi05errata, kazakov04, bruhwiler04}$^{)}$
In this structure, an alkali-metal ion (A) is located in a cage made of O and Os ions. The nearest alkali ions from a particular alkali ion are located in the $\langle 111\rangle$ direction. A large void exists around each alkali ion, and the void becomes larger with decreasing ionic radius of the alkali ion ($r_{\mathrm{A}}$), because the lattice constant ($a$) and the distance between the alkali ion and an oxygen ion ($d_{\mathrm{AO}}$) hardly change with $r_{\mathrm{A}}$ (Table \ref{parameters.t}). The size of the void is characterized by the value $c_{\mathrm{AO}}=d_{\mathrm{AO}}-r_{\mathrm{A}}-r_{\mathrm{O}}$. All three compounds show superconductivity below $T_{\mathrm{c}}=9.6$ K for KOs$_{2}$O$_{6}$, 6.3 K for RbOs$_{2}$O$_{6}$, and 3.3 K for CsOs$_{2}$O$_{6}$. Many studies on their superconductivities have been reported.~\cite{khasanov04, khasanov05, bruhwiler05, magishi05, koda05, kasa06}$^{)}$ The mechanisms of some unusual features such as the nonmonotonic pressure dependence of $T_{\mathrm{c}}$~\cite{mura04, mura05}$^{)}$ and anomalous upper critical fields~\cite{ohmichi06,shiba06}$^{)}$ are still under debate.
\begin{table}[tb]
\caption{Superconducting temperature $T_{\mathrm{c}}$, temperature of first-order phase transition $T_{\mathrm{p}}$, and structural parameters of AOs$_{2}$O$_{6}$.~\cite{hiroi06, yama06, yonephd, shannon76}$^{)}$ }
\label{parameters.t}
\begin{tabular}{ccccccc}
\hline\hline
 					&$T_{\mathrm{c}}$	&$T_{\mathrm{p}}$	&$a$(\AA)		&$d_{\mathrm{AO}}$(\AA)	&$r_{\mathrm{A}}$(\AA)	&$c_{\mathrm{AO}}$(\AA)\\
 \hline
 KOs$_{2}$O$_{6}$		&9.6				&7.5				& 10.09 	&3.13					&1.38	&0.35\\
 RbOs$_{2}$O$_{6}$  	&6.3				&-				& 10.12 	&3.11					&1.52	&0.19\\
 CsOs$_{2}$O$_{6}$  	&3.3				&-				& 10.15 	&3.15					&1.67	&0.08\\
 \hline\hline
\end{tabular}
\end{table}
In all three compounds, a broad peak indicating the contribution from Einstein oscillators was observed in specific-heat experiments,~\cite{hiroi05rbcs, hiroi05errata}$^{)}$ and large atomic displacement parameters (ADP, $U_{\mathrm{iso}}$) of the alkali ions were reported from x-ray diffraction experiments.~\cite{yama06}$^{)}$ These two results are considered as evidence for the existence of a localized mode in AOs$_{2}$O$_{6}$, because the mode would be described as dispersionless in an ideal case and because an alkali ion is likely to vibrate over a wide spatial range due to the weak bonding. In addition to these experiments, a highly anharmonic effective potential for alkali ions was suggested from \textit{ab initio} calculations.~\cite{kunes04,kunes06}$^{)}$ A similar anharmonicity was reported in filled skutterudite and clathrate compounds and is thought to be a characteristic of localized modes.

In addition to the unusual lattice vibrations and exotic superconductivity, an even more interesting and puzzling phenomenon has been observed in KOs$_{2}$O$_{6}$. The specific-heat measurement of KOs$_{2}$O$_{6}$ single crystals has revealed a prominent peak at $T_{\mathrm{p}}= 7.5$ K, which is below $T_{\mathrm{c}}=9.6$ K.~\cite{hiroi05k,hiroi05errata,hiroi06}$^{)}$
A clear jump in the resistivity with a high magnetic field~\cite{hiroi06}$^{)}$ and a kink in the thermal conductivity at the same temperature~\cite{kasa06}$^{)}$ were also reported, whereas no anomaly has been reported for the susceptibility at $T_{\mathrm{p}}$. Recently, an intensity change in the x-ray diffraction upon using a single crystal has been reported.~\cite{yama07}$^{)}$
Since $T_{\mathrm{p}}$ is almost independent of the magnetic field, it is believed that the anomaly does not have a magnetic origin. We now speculate that the localized mode of K in KOs$_{2}$O$_{6}$ plays a significant role in this unassigned phase transition.

In the present study, we have tried to identify the localized mode spectroscopically and to evaluate its anharmonicity using neutron-scattering techniques. In inelastic neutron-scattering (INS) experiments, we have tried to prove the existence of the low-energy dispersionless mode spectroscopically. The energy of the mode was also obtained from INS experiments. In neutron powder-diffraction (NPD) experiments, the $U_{\mathrm{iso}}$ parameter of each ion was determined. To evaluate the anharmonicity of the localized mode, we compared the temperature dependence of the obtained parameters of alkali ions with those of the calculated $U_{\mathrm{iso}}$ values assuming that the alkali ion is well described as a three-dimensional (3D) isotropic harmonic oscillator.
\section{Experimental Procedure}
We typically used 1.1 g of AOs$_{2}$O$_{6}$ powder. KOs$_{2}$O$_{6}$ and CsOs$_{2}$O$_{6}$ were prepared by heating a mixture of A$_{2}$CO$_{3}$ and Os with AgO as an oxidizing agent in two steps via AOsO$_{4}$. RbOs$_{2}$O$_{6}$ was prepared by heating a mixture of Rb$_{2}$O and Os. Powder was placed in an Al cell of 7 mm diameter, and the cell was sealed in an Al can with He gas as a heat-exchange gas.

INS experiments were carried out using a triple-axis spectrometer called PONTA installed by the Institute for Solid State Physics (ISSP), University of Tokyo at the research reactor JRR-3 at the Japan Atomic Energy Agency (JAEA). The final energy of the neutrons was fixed at 14.7 meV using the (002) reflection of pyrolytic graphite (PG). Horizontal collimators were set at 40$^{\prime}$-80$^{\prime}$-radial-blank. Instead of an ordinary flat analyzer,
a horizontally focusing analyzer (HFA) was used to gain more intensity from the localized mode.
Since 5 analyzer crystals were aligned in the HFA, forming a curvature so as to focus the neutrons with 14.7 meV energy to a detector, it gives approximately 5 times larger intensity for $Q$-independent scattering such as incoherent scattering or inelastic scattering from dispersionless modes. The energy resolution under this condition is 1.2 meV at zero energy transfer. A PG filter and a sapphire filter were used to reduce higher-order contamination and the number of fast neutrons.

Applying a so-called incoherent approximation~\cite{bredov67, lynn91}$^{)}$ to the scattering from the powder samples, we can obtain the density of states (DOS) of phonons from INS experiments. Applying the incoherent approximation and neglecting the Debye-Waller factors, the scattering intensity of the one-phonon creation process is written as
\begin{equation}
I \propto \frac{n_{\mathrm{Bose}}(\omega ,T) +1}{\omega} Q^{2} \sum_{j}\frac{\sigma_{\mathrm{coh},j}}{m_{j}} D(\omega),
\end{equation}
where $n_{\mathrm{Bose}}(\omega ,T)$ is the Bose factor, $\sigma_{\mathrm{coh},j}$ is the nuclear coherent cross section of the $j$th atom, $m$ is the mass of the $j$th atom, and $D(\omega)$ is the DOS of the phonons. Note that the values of $\sigma_{\mathrm{coh},j}/m_{j}$ of the relevant nuclei are 0.265 for O, 0.043 for K, 0.074 for Rb, 0.028 for Cs, 0.016 for Os, and 0.076 for W in barn/u and that the scattering intensity from O is the strongest.
In the DOS, a dispersionless mode appears as a delta function. On the other hand, a Debye acoustic mode appears as a constant background at high temperatures or a slope background at low temperatures because   $D(\omega)\propto \omega^{2}$ and $n_{\mathrm{Bose}}(\omega ,T) +1 \propto T/\omega\ (k_{\mathrm{B}}T\gg \hbar \omega)$ or $n_{\mathrm{Bose}}(\omega ,T) +1\simeq1\ (k_{\mathrm{B}}T \ll \hbar \omega)$ for a Debye mode. When $\omega$ reaches the cutoff energy of the acoustic mode, the acoustic mode forms a peak in the DOS due to a van Hove singularity, beyond which the scattering intensity from the mode is zero.

NPD experiments were carried out using a powder diffractometer called HERMES, installed by the Institute for Materials Research (IMR), Tohoku University at JRR-3.~\cite{ohyama98}$^{)}$ The wavelength of the incident neutrons was 1.82646(6)\AA\ using the (331) reflection of Ge. The collimators were set to guide-12$^{\prime}$-blank-18$^{\prime}$. The sample can was mounted on a $^{4}$He closed-cycle refrigerator.
The diffraction patterns were analyzed by the Rietveld method~\cite{riet69}$^{)}$ using the RIETAN-2000 program.~\cite{izumi00}$^{)}$ $U_{\mathrm{iso}}$ appears in the Debye-Waller factor as $\exp(-U_{\mathrm{iso}}Q^{2})$.
\section{Results}
\subsection{Inelastic neutron scattering of AOs$_{2}$O$_{6}$}
\begin{figure}[tb]
\begin{center}
\includegraphics[scale=0.7]{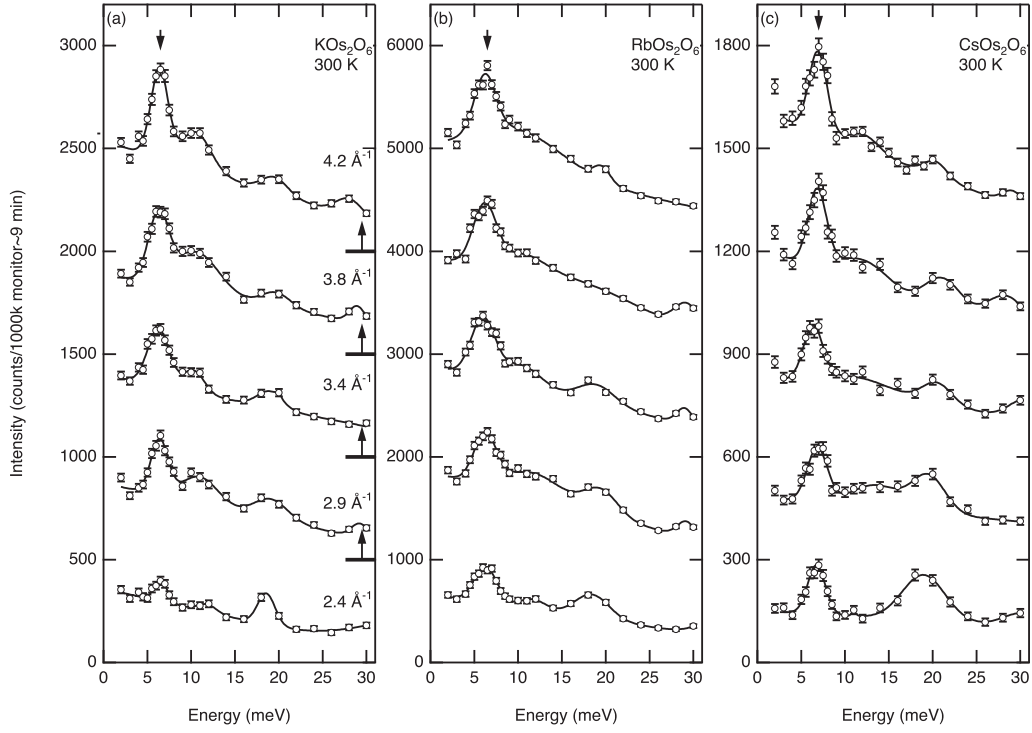}
\end{center}
\caption{Energy dependence of the scattering intensity from AOs$_{2}$O$_{6}$ for different $Q$. (a) A=K, (b) A=Rb, and (c) A=Cs. Each solid line is a fitting result using Gaussians.}

\label{5q.f}
\end{figure}

Figure \ref{5q.f} shows the raw spectra obtained from constant-$Q$ scans of the three samples. In all the samples and for all $Q$, similar profiles were observed: (i) a clear peak around 6.5 meV, (ii) a shoulder around 12 meV, and (iii) another peak around 18 meV.

We first studied the peak around 6.5 meV because possible contributions from Einstein oscillators with an energy of 3.4 meV for KOs$_{2}$O$_{6}$, 5.3 meV for RbOs$_{2}$O$_{6}$, and 6.0 meV for CsOs$_{2}$O$_{6}$ were reported in specific heat experiments.~\cite{hiroi06}$^{)}$

\begin{figure}[tb]
\begin{center}
\includegraphics[scale=1.0]{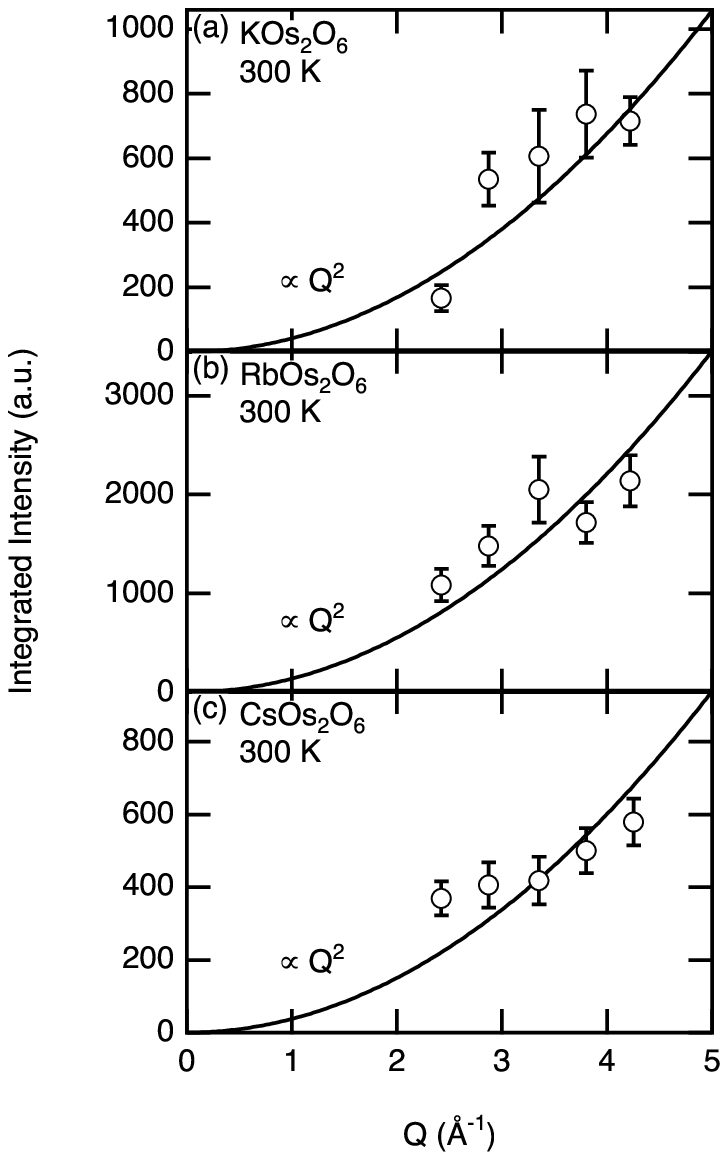}
\end{center}
\caption{Q dependence of the integrated intensities of low-energy peaks. (a) A=K, (b) A=Rb, and (c: A=Cs. The solid lines are the fitting results assuming $I\propto Q^{2}$.}
\label{IvsQ.f}
\end{figure}

\begin{figure}[tb]
\begin{center}
\includegraphics[scale=1.0]{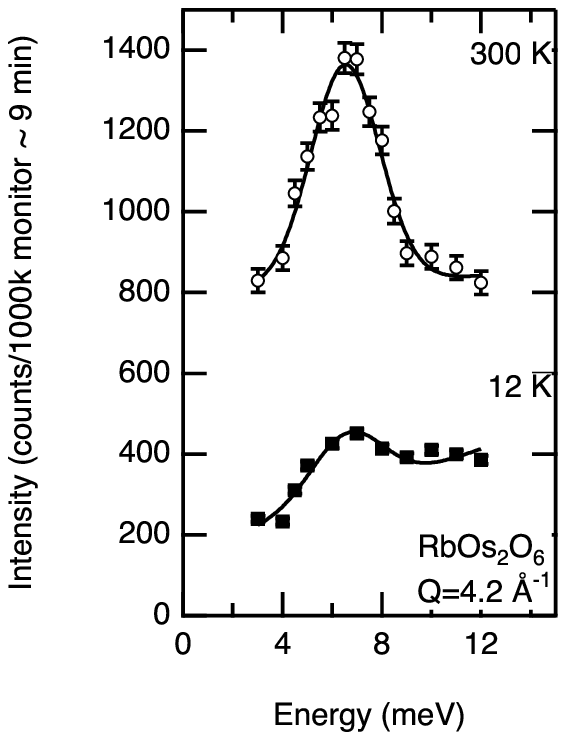}
\end{center}
\caption{Comparison of the scattering intensity of low-energy peaks in RbOs$_{2}$O$_{6}$ at 12 K and room temperature. Each solid line is the fitting result using a Gaussian.}
\label{Rb.f}
\end{figure}

To identify the origin of the first peak, we checked its $Q$ dependence and temperature dependence.  Figures \ref{IvsQ.f} and \ref{Rb.f} show  the $Q$ dependence of the integrated intensities of the first peaks for the three samples and the temperature dependence of the peak for RbOs$_{2}$O$_{6}$, respectively.
The integrated intensity shows monotonic dependence on $Q$ and is roughly proportional to $Q^{2}$. The intensity of the peak from RbOs$_{2}$O$_{6}$ decreases with decreasing temperature and is roughly proportional to $n_{\mathrm{Bose}}(\omega=6.5 \ \mathrm{meV}, T) +1$; the ratios of the scattering intensity at 12 and 300 K were 3.3(4) experimentally and 4.5 theoretically, respectively. From these two sets of dependence, we conclude that the peak originates simply from the lattice dynamics. 
\subsection{Atomic displacement parameters}
\begin{figure}[tb]
\begin{center}
\includegraphics[scale=1.0]{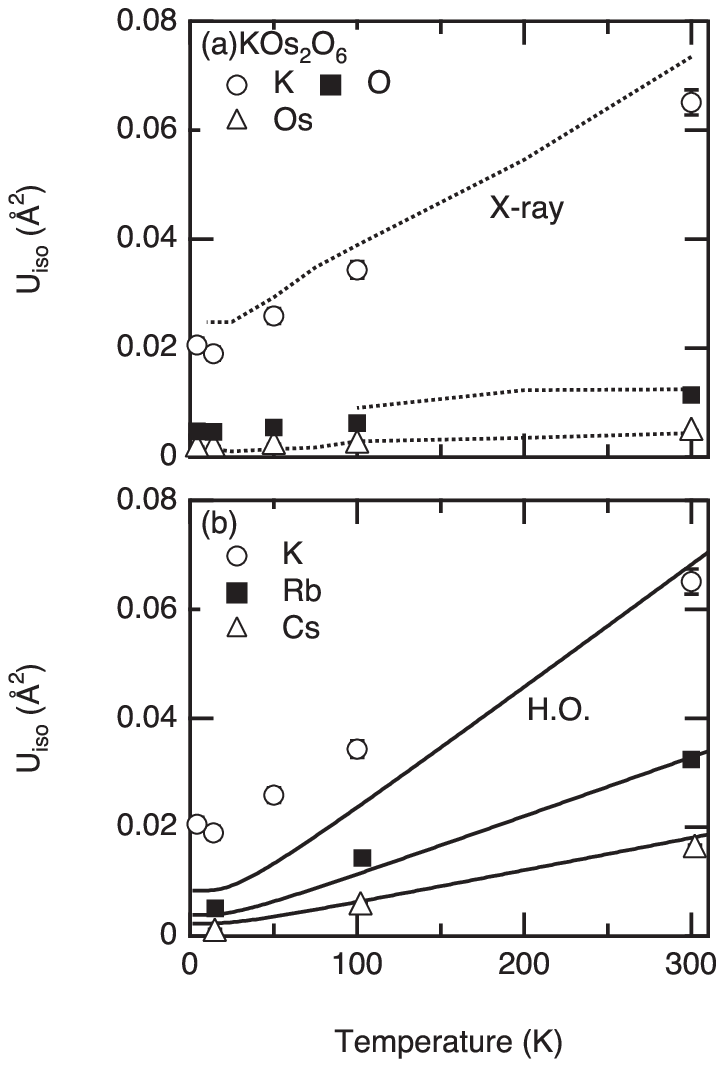}
\end{center}
\caption{Temperature dependence of $U_{\mathrm{iso}}$ for (a) K, Os and O ions in KOs$_{2}$O$_{6}$ and (b) K, Rb, and Cs ions. The dotted lines represents the values obtained from x-ray experiments.~\cite{yama06}$^{)}$ The solid lines are the calculated values for the harmonic oscillators with the energy determined from INS experiments.}
\label{uiso.f}
\end{figure}

We determined $U_{\mathrm{iso}}$ from NPD experiments using the Rietveld method.
In addition to scattering from AOs$_{2}$O$_{6}$, we found diffractions from AOsO$_{4}$, which is obtained in the first reaction process, the starting material Os, and Al, which the cells and cans were made of. In the Rietveld analyses, the angles where Al peaks were expected to appear were ignored. The diffraction profiles were fitted assuming that the main AOs$_{2}$O$_{6}$ phase is $Fd\bar{3}m$ and contains AOsO$_{4}$ and Os as impurities. Free parameters concerning the crystal structures of AOs$_{2}$O$_{6}$ are the lattice constant $a$, the fractional coordinate $x$ of the $48f$ O site, and the three $U_{\mathrm{iso}}$ parameters. The resultant $R$ factors were $R_{\mathrm{wp}}=6.6-7.0$ for KOs$_{2}$O$_{6}$, $5.5-5.8$ for RbOs$_{2}$O$_{6}$, and $7.6-8.9$ for CsOs$_{2}$O$_{6}$ and $R_{\mathrm{e}}=3.9$ for KOs$_{2}$O$_{6}$ and RbOs$_{2}$O$_{6}$, and $4.6-4.8$ for CsOs$_{2}$O$_{6}$. Considering these values of $R$ and the fact that the obtained parameters are very similar to those obtained from x-ray experiments,~\cite{yama06}$^{)}$ we have confirmed that our analysis is reliable. Figure \ref{uiso.f}(a) shows the temperature dependence of the $U_{\mathrm{iso}}$ parameters for the ions in KOs$_{2}$O$_{6}$.
The $U_{\mathrm{iso}}$ of K is much larger than those of Os and O at all temperatures. All the values obtained from the present neutron study are larger than those obtained from previous x-ray studies.~\cite{yama06}$^{)}$ This difference originates from the fact that neutrons are scattered by nuclei with $\delta$-function-like spatial spread, while x-rays are scattered by electrons with a finite spatial spread. Because of these features, neutrons are advantageous in determining $U_{\mathrm{iso}}$.
In Fig. \ref{uiso.f}(b), the $U_{\mathrm{iso}}$ parameters of alkali ions are compared. Upon increasing the ionic radius, $U_{\mathrm{iso}}$ decreases at all temperatures.
\section{Discussion}
\subsection{Origin of the low-lying peak}
\begin{figure}[tb]
\begin{center}
\includegraphics[scale=1]{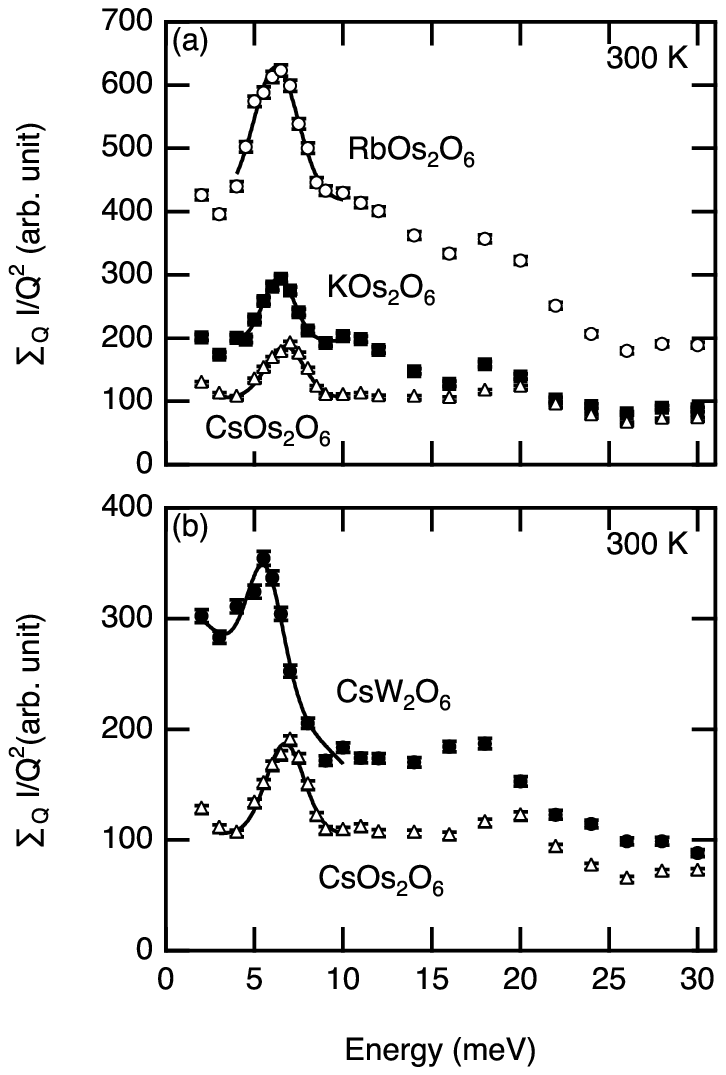}
\end{center}
\caption{Energy dependence of $\sum_QI/Q^{2}$ of (a) AOs$_{2}$O$_{6}$ (open circles: A=K, closed squares: A=Rb, open triangles: A=Cs) and (b) CsB$_{2}$O$_{6}$ (open triangles: B=Os, closed circles: B=W). Each solid line is the fitting result using a Gaussian.}
\label{q2.f}
\end{figure}

To determine which nucleus dominates the scattering, we summed the values of $I(Q)/Q^{2}$. The sum $\sum_QI(Q)/Q^{2}$ should be roughly proportional to the DOS  of phonons with weight $\sigma_{\mathrm{coh},j}/m_{j}$. Figure \ref{q2.f}(a) shows the energy spectra of $\sum_QI(Q)/Q^{2}$ for the three samples.
We first discuss the origin of the low-lying peak obtained from the INS of AOs$_{2}$O$_{6}$ and CsW$_{2}$O$_{6}$. If the peak of  AOs$_{2}$O$_{6}$ originated from an acoustic mode, the intensity of the peak would be proportional to $\sum_{j}\sigma_{\mathrm{coh},j}/m_{j}$ for $j=\ $one A, two Os, and six O, which is almost the same for the three samples. However the intensities of the peak vary significantly among the samples; thus, it is concluded that the low-lying peak is unlikely to originate from the acoustic mode. The ratio of the height of the peak is $1.2:2.6:1$ for $\mathrm{KOs_{2}O_{6}:RbOs_{2}O_{6}:CsOs_{2}O_{6}}$. The respective ratio of the $\sigma_{\mathrm{coh},j}/m_{j}$ of alkali ions multiplied by the effective sample volume determined by the NPD experiments is $1.4:2.8:1$. From this consistency between the two ratios, we conclude that the peak is largely due to scattering by alkali ions. Therefore, the peak around 6.5 meV is dominated by the vibration of alkali ions not the acoustic mode. This is clear evidence for the localized mode of alkali ions. To estimate the energy and width of the observed peak, we fitted the peak with a Gaussian. The fitting results are summarized in Table \ref{fitting.t}. The energies of the peak have almost the same value for the three samples. The widths are somewhat different. Either anharmonicity, hybridization with acoustic modes, or the dispersion of the mode itself can broaden the peak. However, we cannot determine the cause only from these INS powder experiments.
\begin{table}[tb]
\caption{Energy ($E_{\mathrm{R}}$) and  full width at half maximum (FWHM, $W_{\mathrm{R}}$) of the observed peak in AOs$_{2}$O$_{6}$ (unit: meV).}
\label{fitting.t}
\begin{tabular}{ccc}
\hline\hline
 					&$E_{\mathrm{R}}$	&$W_{\mathrm{R}}$	\\
 \hline
 KOs$_{2}$O$_{6}$		&6.4(1)			&1.2(3)			\\
 RbOs$_{2}$O$_{6}$  	&6.5(1)			&2.8(2)			\\
 CsOs$_{2}$O$_{6}$  	&6.8(1)			&2.0(3)			\\
 \hline\hline
\end{tabular}
\end{table}
The shoulder around 12 meV is attributed to the van Hove singularity of the acoustic modes (and possibly other optical modes) because an acoustic mode forms a constant background at high temperatures before it reaches the cutoff energy. The other peak around 18 meV is attributed to other optical modes.
\subsection{$U_{\mathrm{iso}}$ parameters and anharmonicity}
We carefully examined the $U_{\mathrm{iso}}$ parameters determined from NPD. In ordinary materials, $U_{\mathrm{iso}}$ exhibit behaviors expected from the Debye mode. To check that an alkali ion is a harmonic oscillator in AOs$_{2}$O$_{6}$, we tried to determine whether $U_{\mathrm{iso}}$ is attributed to the Einstein mode. In this examination, we compared $U_{\mathrm{iso}}$ obtained from NPD experiments with that obtained from the 3D harmonic oscillator. $U_{\mathrm{iso}}$ of the 3D harmonic oscillator is given by 
\begin{equation}
U_{\mathrm{iso}}=\frac{\hbar}{2 m \omega _{\mathrm{E}}}\coth \left( \frac{\hbar \omega}{2 k_{\mathrm{B}}T}\right), 
\end{equation}
where $m$ and $\omega _{\mathrm{E}}$ are the mass and energy of the oscillator, respectively. We adopted the peak energy in INS experiments as the energy of the 3D isotropic harmonic oscillator. The experimentally determined  $U_{\mathrm{iso}}$ parameters of Rb and Cs agree well with the calculated values, while the experimentally obtained $U_{\mathrm{iso}}$ parameters of K are larger than the calculated values, particularly at low temperatures. This discrepancy is not resolved even if we add a temperature-independent constant to $U_{\mathrm{iso}}$, which is often used to describe static disorder. The existence of another lower mode in the energy suggested by the specific-heat experiments also would not resolve the discrepancy because such a mode only alters the slope of $U_{\mathrm{iso}}$ at high temperatures. We thus conclude that this discrepancy is largely due to anharmonicity of the effective potential for K ions.

\subsection{Energy of the localized mode}
The energies of the localized modes in the three AOs$_{2}$O$_{6}$ (A=K, Rb, Cs) samples have almost the same values.
Assuming that the localized mode is an assembly of isotropic harmonic oscillators, its energy is described as $E \propto \sqrt{k/m_{\mathrm{A}}}$, where $k$ is a spring constant. The ratio of $k$ has been determined to be $1:1.6:4.1$ for $\mathrm{K:Rb:Cs}$ from INS experiments.
Thus, we speculate that the larger the alkali ion, the stronger the bonding between the alkali ion and its cage. However, since the anharmonicity of the potential for K ions is substantial, we must consider the effect of the anharmonicity of the mode on the energy in future discussion.

The energy of the localized mode determined in the present study agrees well with the Einstein temperature determined by the specific-heat measurements for RbOs$_{2}$O$_{6}$ and CsOs$_{2}$O$_{6}$ but differs substantially for KOs$_{2}$O$_{6}$. We speculate that this discrepancy is possibly due to the anharmonicity of the mode.~\cite{hiroi05errata, hiroi05rbcs}$^{)}$
\subsection{Effects of the cage}
We also performed INS experiments on CsW$_{2}$O$_{6}$, where the cage consists of not Os but W.~\cite{cava93}$^{)}$ Figure \ref{q2.f}(b) shows a comparison of $\sum_QI(Q)/Q^{2}$ for CsOs$_{2}$O$_{6}$ and CsW$_{2}$O$_{6}$. A clear low-energy peak was also observed in CsW$_{2}$O$_{6}$, but the energy is 5.6 meV, considerably different from that of CsOs$_{2}$O$_{6}$. The shape of the background is also different.
The difference between the energy of the peak in CsOs$_{2}$O$_{6}$ and that in CsW$_{2}$O$_{6}$ indicates that the atoms forming the cage have a significant effect on the localized mode.  One possible explanation for this is that an electron-phonon interaction contributes to the localized mode through the charge screening of the alkali ion by the cage.
\section{Conclusions}
We successfully observed the localized modes of the alkali ion in AOs$_{2}$O$_{6}$ (A=K, Rb, Cs) and CsW$_{2}$O$_{6}$ from INS experiments. It was found that the energy of the mode weakly depends on the size of the alkali metal ion and strongly depends on the atom constituting the cage. The motion of Rb and Cs seems to be harmonic, while that of K deviates from the harmonic oscillator, particularly at low temperature.
%
%
%
%
%
%
%
%
%
\section*{Acknowledgments}
We are grateful to Dr. Matsuura for his help in INS experiments, to Dr. Nishi for his advice on the use of the HFA, to Professor Ohoyama for the use of HERMES, to Professor Takigawa, Professor Yamamuro, and Dr. Tsutsui for useful discussions on data interpretation, and to Professor Sato for his advice on data analysis. Neutron-scattering experiments at JRR-3 were carried out under the ISSP user program.
%
%

\end{document}